\documentclass[12pt]{article}
\usepackage{epsfig}
\usepackage{wrapfig}
\usepackage{subfigure}
\usepackage{amsmath}
\usepackage{hhline}
\usepackage{amssymb}
\usepackage{times}
\usepackage{cite}

\newlength{\dinwidth}
\newlength{\dinmargin}
\begin{document}  
\newcommand{\be}{\begin{equation}}
\newcommand{\ee}{\end{equation}}
\newcommand{\bea}{\begin{eqnarray}}
\newcommand{\eea}{\end{eqnarray}}
\newcommand{\nn}{\nonumber}
\newcommand{\sr}{\stackrel}
\newcommand{\D}{\displaystyle}
\newcommand{\half}{\textstyle{\frac{1}{2}}}
\newcommand{\bs}{\boldsymbol}

\newcommand{\pom}{{I\!\!P}}
\newcommand{\reg}{{I\!\!R}}
\newcommand{\gap}{\stackrel{>}{\sim}}
\newcommand{\lap}{\stackrel{<}{\sim}}

\newcommand{\alps}{\alpha_s}
\newcommand{\sqrts}{$\sqrt{s}$}
\newcommand{\LO}{$O(\alpha_s^0)$}
\newcommand{\Oa}{$O(\alpha_s)$}
\newcommand{\Oaa}{$O(\alpha_s^2)$}
\newcommand{\PT}{p_{\perp}}
\newcommand{\PO}{I\!\!P}
\newcommand{\xpomlo}{3\times10^{-4}}  
\newcommand{\dgr}{^\circ}
\newcommand{\pbarnt}{\,\mbox{{\rm pb$^{-1}$}}}

%
%
\newcommand{\tg}{\theta_{\gamma}}
\newcommand{\te}{\theta_e}
%
%
\newcommand{\qsq}{\mbox{$Q^2$}}
\newcommand{\Qsq}{\mbox{$Q^2$}}
\newcommand{\s}{\mbox{$s$}}
\newcommand{\ttra}{\mbox{$t$}}
\newcommand{\modt}{\mbox{$|t|$}}
\newcommand{\eminpz}{\mbox{$E-p_z$}}
\newcommand{\eminpzs}{\mbox{$\Sigma(E-p_z)$}}
\newcommand{\rap}{\ensuremath{\eta^*} }
\newcommand{\W}{\mbox{$W$}}
\newcommand{\w}{\mbox{$W$}}
\newcommand{\Q}{\mbox{$Q$}}
\newcommand{\q}{\mbox{$Q$}}
\newcommand{\xB}{\mbox{$x$}}  
\newcommand{\xF}{\mbox{$x_F$}}  
\newcommand{\xg}{\mbox{$x_g$}}  
\newcommand{\xbj}{x}
\newcommand{\xpom}{x_{\PO}}
\newcommand{\y}{\mbox{$y~$}}
\newcommand{\gp}{\ensuremath{\gamma^*}p }
\newcommand{\gammasp}{\ensuremath{\gamma}*p }
\newcommand{\gammap}{\ensuremath{\gamma}p }
\newcommand{\gsp}{\ensuremath{\gamma^*}p }
\newcommand{\epem}{\mbox{$e^+e^-$}}
\newcommand{\ep}{\mbox{$ep~$}}
\newcommand{\epl}{\mbox{$e^{+}$}}
\newcommand{\emi}{\mbox{$e^{-}$}}
\newcommand{\epm}{\mbox{$e^{\pm}$}}
%
\newcommand{\photon}{\mbox{$\gamma$}}
\newcommand{\phib}{\mbox{$\varphi$}}
\newcommand{\rh}{\mbox{$\rho$}}
\newcommand{\rhz}{\mbox{$\rh^0$}}
\newcommand{\ph}{\mbox{$\phi$}}
\newcommand{\om}{\mbox{$\omega$}}
\newcommand{\ome}{\mbox{$\omega$}}
\newcommand{\jpsi}{\mbox{$J/\psi$}}
\newcommand{\JPSI}{J/\psi}
\newcommand{\ups}{\mbox{$\Upsilon$}}
\newcommand{\bsl}{\mbox{$b$}}
%
%
\newcommand{\cm}{\mbox{\rm cm}}
\newcommand{\GeV}{\mbox{\rm GeV}}
\newcommand{\gev}{\mbox{\rm GeV}}
\newcommand{\GeVx}{\rm GeV}
\newcommand{\gevx}{\rm GeV}
\newcommand{\MeV}{\mbox{\rm MeV}}
\newcommand{\mev}{\mbox{\rm MeV}}
\newcommand{\MeVx}{\mbox{\rm MeV}}
\newcommand{\mevx}{\mbox{\rm MeV}}
\newcommand{\GeVsq}{\mbox{${\rm GeV}^2$}}
\newcommand{\gevsq}{\mbox{${\rm GeV}^2$}}
\newcommand{\gevsqc}{\mbox{${\rm GeV^2/c^4}$}}
\newcommand{\gevcsq}{\mbox{${\rm GeV/c^2}$}}
\newcommand{\mevcsq}{\mbox{${\rm MeV/c^2}$}}
\newcommand{\GeVsqm}{\mbox{${\rm GeV}^{-2}$}}
\newcommand{\gevsqm}{\mbox{${\rm GeV}^{-2}$}}
\newcommand{\nb}{\mbox{${\rm nb}$}}
\newcommand{\nbinv}{\mbox{${\rm nb^{-1}}$}}
\newcommand{\pbinv}{\mbox{${\rm pb^{-1}}$}}
\newcommand{\mm}{\mbox{$\cdot 10^{-2}$}}
\newcommand{\mmm}{\mbox{$\cdot 10^{-3}$}}
\newcommand{\mmmm}{\mbox{$\cdot 10^{-4}$}}
\newcommand{\degr}{\mbox{$^{\circ}$}}
%
%
\def\gsim{\,\lower.25ex\hbox{$\scriptstyle\sim$}\kern-1.30ex%
  \raise 0.55ex\hbox{$\scriptstyle >$}\,}
\def\lsim{\,\lower.25ex\hbox{$\scriptstyle\sim$}\kern-1.30ex%
  \raise 0.55ex\hbox{$\scriptstyle <$}\,}
%
%


\begin{titlepage}

\noindent
\begin{flushleft}
{\tt DESY 09-109    \hfill    ISSN 0418-9833} \\
{\tt July 2009}                  \\
\end{flushleft}

\vspace{2cm}
\begin{center}
\begin{Large}

{\bf  
Deeply Virtual Compton Scattering and its Beam Charge Asymmetry in \begin{boldmath}$e^\pm p$\end{boldmath}
Collisions at HERA
}

\vspace{2cm}

H1 Collaboration

\end{Large}
\end{center}

\vspace{2cm}

\begin{abstract}

A measurement of elastic deeply virtual Compton 
scattering $\gamma^* p \rightarrow \photon p$ using $e^+ p$ and $e^- p$ collision data
recorded with the H1 detector at HERA is presented. 
The analysed data sample corresponds to 
an integrated luminosity of $306$~pb$^{-1}$, almost equally shared between both beam charges.
The cross section is measured as a function of the virtuality
$Q^2$ of the exchanged photon
and the centre-of-mass energy $W$ of the $\gamma^*p$ system 
in the kinematic domain 
$6.5 < Q^2 < 80$~GeV$^2$, $30 < W < 140$~GeV and $|t| < 1$~GeV$^2$,
where  $t$ denotes the  squared momentum transfer at the proton vertex.
The cross section is determined differentially in $t$ 
for different $Q^2$ and $W$ values and
exponential $t$-slope parameters are derived. 
Using $e^+ p$ and $e^- p$ data samples, a beam charge asymmetry is extracted for the first
time in the low Bjorken $x$ kinematic domain.
The observed asymmetry is attributed to the interference between Bethe-Heitler  and deeply virtual Compton scattering processes. 
Experimental results are discussed in the context of two different models, one based on generalised parton distributions and one based on the dipole
approach.

\end{abstract}

\vspace{1.5cm}

\begin{center}
Accepted by Phys. Lett. {\bf B}
\end{center}

\end{titlepage}

\begin{flushleft}

F.D.~Aaron$^{5,49}$,           
M.~Aldaya~Martin$^{11}$,       
C.~Alexa$^{5}$,                
K.~Alimujiang$^{11}$,          
V.~Andreev$^{25}$,             
B.~Antunovic$^{11}$,           
S.~Backovic$^{30}$,            
A.~Baghdasaryan$^{38}$,        
E.~Barrelet$^{29}$,            
W.~Bartel$^{11}$
% ...

\bigskip{\it
 $ ^{1}$ I. Physikalisches Institut der RWTH, Aachen, Germany \\
 $ ^{2}$ Vinca  Institute of Nuclear Sciences, Belgrade, Serbia \\
% ...
$ ^{5}$ National Institute for Physics and Nuclear Engineering (NIPNE) ,
          Bucharest, Romania \\
% ...
$ ^{11}$ DESY, Hamburg, Germany \\
% ...
$ ^{25}$ Lebedev Physical Institute, Moscow, Russia$^{ e}$ \\
% ...
$ ^{29}$ LPNHE, Universit\'{e}s Paris VI and VII, IN2P3-CNRS,
          Paris, France \\
 $ ^{30}$ Faculty of Science, University of Montenegro,
          Podgorica, Montenegro$^{ e}$ \\
% ...
$ ^{38}$ Yerevan Physics Institute, Yerevan, Armenia \\
% ...
$ ^{49}$ Also at Faculty of Physics, University of Bucharest,
          Bucharest, Romania \\
% ...
$ ^e$ Supported by the Deutsche Forschungsgemeinschaft \\
% ...
}
\end{flushleft}

\newpage

\section{Introduction}
\label{sec:intro}

Measurements of the deep-inelastic scattering (DIS) of leptons and nucleons allow the
extraction of Parton Distribution Functions (PDFs). While
these functions provide crucial input to perturbative Quantum Chromodynamic (QCD) calculations,
they do not provide a complete picture of the partonic
structure of nucleons. In particular, PDFs contain neither information on the correlations between
partons nor on their transverse spatial distribution.

Hard exclusive particle production, without excitation or dissociation of the nucleon, have emerged in recent years as prime candidates to address these issues~\cite{bernard,bel,buk,diehl,strik,muller,diehl3}.
Among them, deeply virtual Compton scattering (DVCS) on the proton ($\gamma^*p\to\gamma p$) is the simplest.
The DVCS reaction can be regarded as the elastic scattering of the
virtual photon off the proton via a colourless exchange, producing a real photon in the final state. 
In the Bjorken scaling regime, corresponding to large virtuality  $Q^2$ of the exchanged photon and $|t|/Q^2 \ll 1$, where  $t$ is the  squared momentum transfer at the proton vertex, QCD calculations assume that the exchange involves two partons in a colourless configuration, having different longitudinal and transverse momenta. 
These unequal momenta, or skewing, are a consequence of the mass
difference between the incoming virtual photon and the outgoing real
photon and may
 be interpreted in the context of generalised parton distributions (GPDs) or dipole amplitudes, respectively. 
In basic terms, a GPD (off-diagonal parton distribution)
is the transition amplitude for removing a parton from the
fast moving proton and reabsorbing it with a different momentum,
thereby imparting a certain momentum transfer to the proton.
In the dipole approach the virtual photon fluctuates into a colour singlet $q {\bar q}$ pair (or dipole) of a transverse size
$r \sim 1/Q$, which subsequently undergoes hard scattering with the gluons in the proton.
The $t$-dependence of the DVCS cross section carries information on the
transverse momentum of partons.

In the kinematic range of the HERA collider, where DVCS is accessed through the reaction $e^\pm  p \rightarrow e^\pm  \photon  p$~\cite{dvcsh1,Adloff:2001cn,dvcsh1a,dvcszeus,dvcszeusb}, the DVCS amplitude is mainly imaginary~\cite{bel}, while the change of
the amplitude with energy gives rise to a small real part. 
This reaction also receives a contribution from the purely 
electromagnetic Bethe-Heitler (BH) process, where the photon is emitted from the electron.
The interference between DVCS and BH processes allows the extraction of the real part of the amplitude.
In addition, the real part of the DVCS amplitude can
 be related to its imaginary part using dispersion relations. 
In the high energy limit at low momentum fraction $x$, the dispersion relations take a simple form~\cite{Hebecker:2000xs}  which can therefore be used for the DVCS process to verify the consistency between measurements of the real and imaginary parts of the amplitude.

This paper presents a measurement of DVCS cross sections as a function of $Q^2$ and the $\gamma^*p$ centre-of-mass energy $W$.
The single differential cross section $d\sigma / dt$ is also extracted.
The data were recorded with the H1 detector in the years $2004$ to $2007$,
during which period HERA collided protons of $920$~GeV energy with
$27.6$~GeV electrons and positrons.
The total integrated luminosity of the data is $306$~pb$^{-1}$.
The data comprise $162$~pb$^{-1}$ recorded in $e^+p$ and $144$ pb$^{-1}$ in $e^-p$ collisions.
During this HERA~II running period, the electron\footnote{
  In this paper the term ``electron'' is used generically to refer to both electrons and positrons, unless otherwise stated.} beam was longitudinally polarised, at a level of typically $35\%$. For this analysis, the periods with left-handed and right-handed beams are combined and the analysed data samples have a left-handed residual polarisation of $1\%$ and $5\%$ for $e^+p$ and $e^-p$ collisions, respectively.
Cross section measurements are carried out 
in the kinematic range 
$6.5 < Q^2 < 80$~GeV$^2$, $30 < W < 140$~GeV 
and $|t| <$ 1 GeV$^2$.
The range in $x \simeq Q^2/W^2$ of the present measurement extends from $5 \cdot 10^{-4}$ to
 $10^{-2}$.
The cross section measurements of this analysis supersede those of a previous H1 publication~\cite{dvcsh1}, in which less than half of the present HERA II data was used.
It is complementary to measurements performed at lower $Q^2$ using HERA I data~\cite{dvcsh1a}.
In addition, using both beam charges, the beam charge asymmetry of the interference between the BH and DVCS processes 
is measured for the first time at a collider. 
%
%

\section{Theoretical Framework}
\label{sec:theo}

In this paper, cross section measurements are compared to predictions based either on GPDs or on a dipole approach.
At the present level of understanding, the pure GPD approach and dipole models,
based on the proton-dipole amplitude,
are not connected. 
However, in the low $x$ domain, dipole amplitudes could be used to provide 
parameterisations for GPDs at a certain scale~\cite{McDermott:2001pt}.
In this context, the DVCS process is interesting as calculations are simplified by the absence of an unknown vector meson wave function.
The GPD model~\cite{muller} used here has been shown to describe previous data. 
It is based on partial wave expansions of DVCS amplitudes and
is a first attempt to parametrise all GPDs over the full kinematic domain.
The  dipole model~\cite{gregory}, with a limited number of parameters, describes a large panel of low $x$ measurements at HERA, from inclusive 
to exclusive processes. 
In this model, mainly using the gluon density extracted from fits to $F_2$ data,
the DVCS cross section is  computed using a universal dipole amplitude.

For GPD models, a direct measurement of the real part of the DVCS amplitude is an important issue, as it gives an increased sensitivity to the parameterisation of the GPDs~\cite{bel,muller}.
Indeed, a calculation of the real part of the DVCS amplitude requires a parametrisation of the GPDs
over the full $x$ range. 
Considering the large flexibility in the parameterisation of the GPDs,  this is an 
important quantity to qualify the correct approach with GPDs.
In the dipole approach, as the dipole amplitude refers only to the imaginary part, the magnitude of the real part can be predicted using a dispersion relation.

In high energy electron-proton collisions at HERA,
DVCS and BH processes contribute to the reaction 
$e^\pm  p \rightarrow e^\pm  \photon  p$.
%
The BH cross section is precisely calculable in QED.
Since these two processes have an identical final state, they interfere.
The squared photon production amplitude is then given by 

\begin{equation} \label {eqn:tau}
\left| A \right|^2 
= \left| A_{{\scriptscriptstyle BH}} \right|^2 + 
\left| A_{{\scriptscriptstyle DVCS}} \right|^2 + \underbrace{
A_{{\scriptscriptstyle DVCS}} \, A_{{\scriptscriptstyle BH}}^* 
+ A_{{\scriptscriptstyle DVCS}}^* \, A_{{\scriptscriptstyle BH}}}_I,
\end{equation}

\noindent where $A_{\scriptscriptstyle BH}$ is the BH amplitude,
$A_{\scriptscriptstyle DVCS}$ represents the DVCS amplitude and
$I$ denotes the interference term.
In the leading twist approximation, the interference term can be written
quite generally as a linear combination of harmonics of the azimuthal angle
$\phi$. As defined in~\cite{bel}, $\phi$ is the angle between the 
plane containing the incoming and outgoing leptons 
and the plane formed by the virtual and real photons. 
For an unpolarised proton beam and if only the first harmonic in $\cos\phi$ and $\sin\phi$, which are dominant at low $x$~\cite{muller}, are considered, 
the interference term $I$ can be written as

\begin{equation}
I \propto 
-C \, 
[  a_1 \cos \phi \, \mathrm{Re}  A_{DVCS}  
+ a_2 P_l \sin \phi \, \mathrm{Im}  A_{DVCS}  
],
\end{equation}

\noindent where $C = \pm 1$ is the charge of the lepton beam, $P_l$
its longitudinal polarisation
and $a_1$ and $a_2$ are 
functions of the ratio of longitudinal to 
transverse virtual photon flux~\cite{bernard,bel,buk,diehl,strik,muller}.
Cross section measurements which are integrated over $\phi$ are 
not sensitive 
to the interference term.
The measurement of the cross section asymmetry with respect to the beam
charge as a function of $\phi$ allows to access the interference term.
The beam charge asymmetry (BCA) of the cross section is defined as

\begin{equation}
A_C(\phi) = \frac{d\sigma^+/d\phi -d\sigma^-/d\phi} 
{d\sigma^+/d\phi + d\sigma^-/d\phi},
\label{bcadef}
\end{equation} 

\noindent where  $d\sigma^+/d\phi$ and $d\sigma^-/d\phi$ are the differential $ep \rightarrow ep\gamma$ cross sections measured in $e^{+}p$ and $e^{-}p$ collisions, respectively.

Considering the low residual polarisation of the data and the theoretical expression of $a_1$ and $a_2$~\cite{bel},  $a_1 \gg a_2 P_l$ and the contribution of the $\sin \phi$ term is neglected. 
Therefore, $A_C(\phi)$ can be expressed as 

\begin{equation}
A_C(\phi) = p_1 \cos \phi = 2 A_{BH} \frac{\mathrm{Re}  A_{DVCS} } 
{ |A_{DVCS}|^2+ |A_{BH}|^2 } \cos \phi.
\label{eq1}
\end{equation}

\noindent The term $|A_{DVCS}|^2$ can be derived directly from the DVCS cross 
section measurement $\sigma_{DVCS} = |A_{DVCS}^2| / (16 \pi b)$, where $b$ is 
the slope of the exponential $t$-dependence $e^{-b|t|}$ of the DVCS cross
section.
As the BH amplitude is precisely known, the measured asymmetry is directly proportional to the real part of the DVCS amplitude and the ratio between real and imaginary parts of the DVCS amplitude, $\rho=\mathrm{Re}  A_{DVCS} / \mathrm{Im}  A_{DVCS}$, can be extracted.
This ratio $\rho$ can also be derived using a dispersion relation~\cite{muller,diehl2}. 
In the high energy limit, at low $x$ and when the $W$ dependence of the cross section
is parameterised by a single term $W^{\delta(Q^2)}$, the dispersion relation can be written as~\cite{Hebecker:2000xs}

\begin{equation}
\rho=\mathrm{Re}  A_{DVCS} / \mathrm{Im}  A_{DVCS} 
= \tan \left( \frac{\pi \delta(Q^2)} {8} \right).
\label{dispersion}
\end{equation}

The ratio $\rho$ can therefore be determined directly from the energy dependence of the DVCS cross section parameterised by $\delta(Q^2)$. 
Comparison between the $\rho$ values calculated from the energy dependence of the DVCS amplitude and from its real part therefore provides an important consistency test of the measured BCA.

\section{Experimental Conditions and Monte Carlo Simulation} \label{simul}

A detailed description of the H1 detector can be found in~\cite{h1dect}.
Here, only the detector components relevant for the present analysis are
described. 
H1 uses a right-handed coordinate system with the $z$ axis along
the beam direction, the $+z$ or ``forward'' direction being that of the outgoing proton beam.
The polar angle $\theta$ is defined with respect to the $z$ axis and the
pseudo-rapidity is given by $\eta=-\ln \tan \theta /2$. 

The SpaCal~\cite{Appuhn:1996na}, a lead scintillating fibre calorimeter, 
covers the backward 
region ($153 ^{\rm \circ} < \theta < 176 ^{\rm \circ}$).
Its energy resolution for electromagnetic showers is $\sigma(E)/E
\simeq 7.1\%/\sqrt{E/{\rm GeV}} \oplus 1\%$. 
The liquid argon (LAr) calorimeter ($4^{\rm \circ} \leq \theta \leq
154^{\rm \circ}$) is situated inside a solenoidal magnet. 
The energy resolution for electromagnetic showers is 
$\sigma(E)/E \simeq 11\%/\sqrt{E/{\rm GeV}}$ as obtained from test beam 
measurements~\cite{Andrieu:1994yn}.
The main component of the central tracking detector is the central jet
chamber CJC ($20^{\rm \circ} < \theta < 160^{\rm \circ}$) which consists of 
two coaxial cylindrical drift chambers
with wires parallel to the beam direction.
The measurement of charged particle transverse momenta is performed
in the magnetic field of $1.16$~T, with a resolution of $\sigma_{P_T}/P_T = 0.002 P_T / \rm{GeV} \oplus 0.015$.
The innermost proportional chamber CIP~\cite{CIP} ($9^\circ < \theta < 171^\circ$)
is used in this analysis to complement the CJC in the backward region for the reconstruction of the interaction vertex.
The forward muon detector (FMD) consists of
a series of drift chambers covering the range $1.9<\eta<3.7$. 
Primary particles produced at larger $\eta$ can be detected indirectly in the FMD if they undergo a secondary scattering with the beam pipe or other adjacent material. 
Therefore, the FMD is used in this analysis to provide an additional veto against inelastic or proton dissociative events.
The luminosity is determined from the rate of Bethe-Heitler processes
measured using a calorimeter located
close to the beam pipe at $z=-103~{\rm m}$ in the backward direction.

A dedicated event trigger was set up for this analysis.
It is based on topological and neural network algorithms
and uses correlations between electromagnetic energy 
deposits of electrons or photons in both the LAr and the SpaCal \cite{roland}.
The combined trigger efficiency is $98$\%.

Monte Carlo (MC) simulations are used to estimate the background contributions and the corrections for the QED radiative effects and for the finite acceptance and the resolution of the detectors.
Elastic DVCS events in $ep$ collisions are generated using the Monte Carlo 
generator MILOU~\cite{milou}, based on the cross section calculation 
from~\cite{ffspaper} and using a $t$-slope parameter $b=5.4$~GeV$^{-2}$, 
as measured in this analysis (see section~\ref{sec:exp_xsec}). The photon 
flux is taken from~\cite{flux}.
Inelastic DVCS events in which the proton dissociates into a baryonic system $Y$ are
also simulated with MILOU setting the $t$-slope
$b_{inel}$ to $1.5$~GeV$^{-2}$, as determined in a dedicated study (see section~\ref{sec:pdiss}).
The Monte Carlo program COMPTON~2.0~\cite{compton2} is used to
simulate elastic and inelastic BH events.
In the generated MC events, no interference between DVCS and BH processes is included.
Background from
diffractive  meson events is simulated using the
DIFFVM MC generator~\cite{diffvm}.
All generated events are passed through a detailed, GEANT~\cite{Brun:1987ma} based simulation of the H1
detector and are subject to the same reconstruction and analysis chain
as are the data.

\section{Event Selection}  \label{selection}
%
%
In elastic DVCS events, the scattered electron and the photon
are the only particles that are expected to give signals in the detector.
The scattered proton escapes undetected through the beam pipe.
The selection of the analysis event sample requires a scattered electron 
and a photon identified as compact and isolated electromagnetic showers in the SpaCal and in the LAr, respectively.
The electron candidate is required to have an energy above $15$ GeV.
The photon is required to have a transverse momentum $P_T$ above $2$ GeV 
and a polar angle between $25^{\circ}$ and $145^\circ$.
Events are selected if there are either no tracks at all or a single central track which is associated with the scattered electron.
In order to reject inelastic and proton dissociation events,
no further energy deposit in the LAr calorimeter larger than
$0.8$~GeV is allowed and no activity above
the noise level should be present in the FMD.
The influence of QED radiative corrections is reduced by the requirement that
the longitudinal momentum balance $E - P_z$ be greater than $45$~GeV.
Here, $E$ denotes the energy and $P_z$ the momentum along the beam axis of all measured final state particles.
To enhance the DVCS signal with respect to the BH
contribution and to ensure a large acceptance, the
kinematic domain is  restricted to
$6.5<Q^2<80 $ GeV$^2$ and
$30<W<140 $~GeV.

The reconstruction method for the kinematic variables $Q^2$, $x$ and 
$W$ relies on the measured polar angles of the final 
state electron and photon (double angle method)~\cite{dvcsh1}.
The variable $t$ is approximated by the negative square of the 
transverse momentum of the outgoing proton, computed from the vector sum of the transverse momenta of the final state photon and the scattered electron.
The resolution of the $t$ reconstruction varies from $0.06$ at low $|t|$ to $0.20$~GeV$^2$ at high $|t|$.

\begin{figure}[!htbp]
\vspace*{2cm}
\begin{center}
 \includegraphics[totalheight=5.7cm]{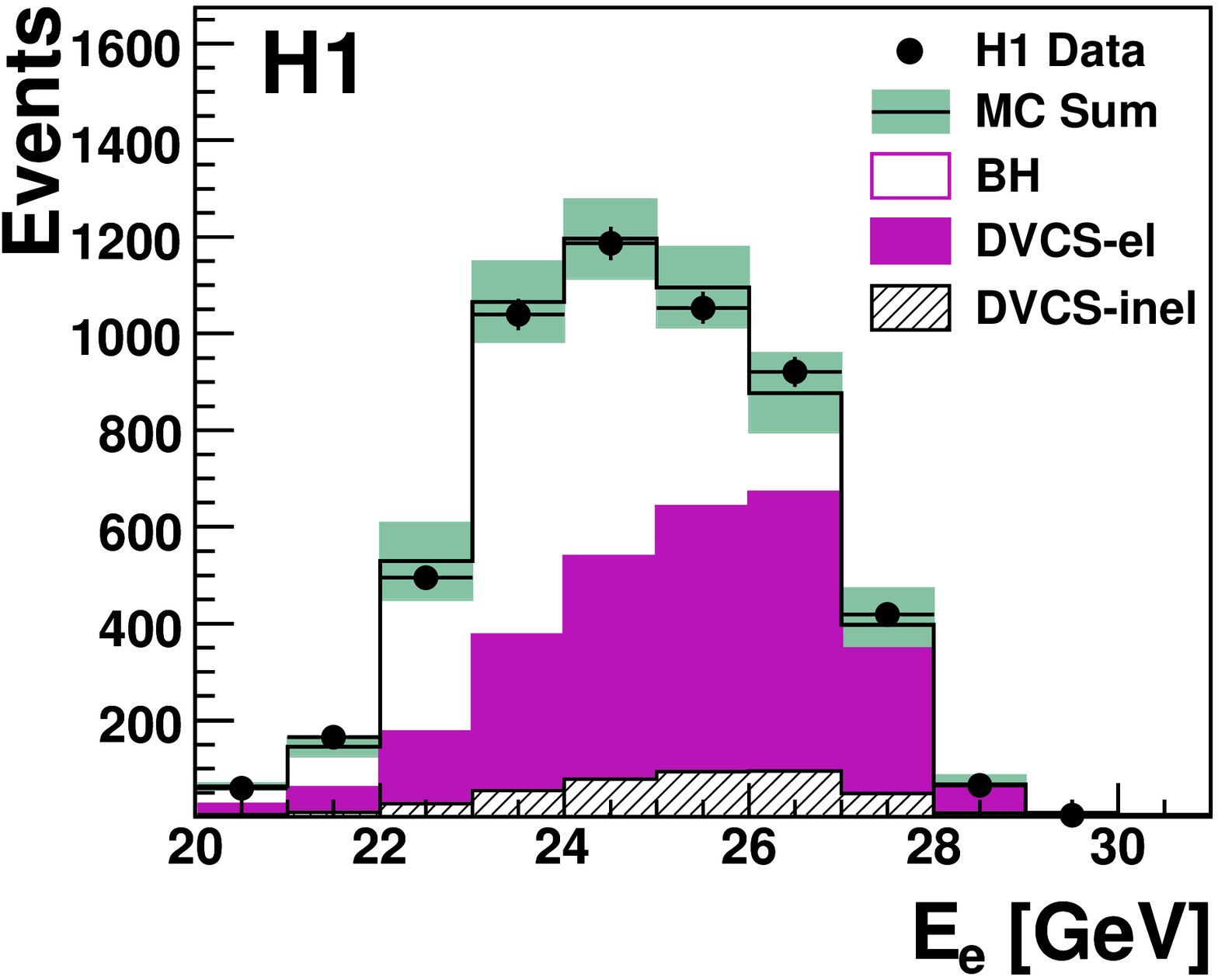}\put(-55,40){{(a)}}
 \includegraphics[totalheight=5.7cm]{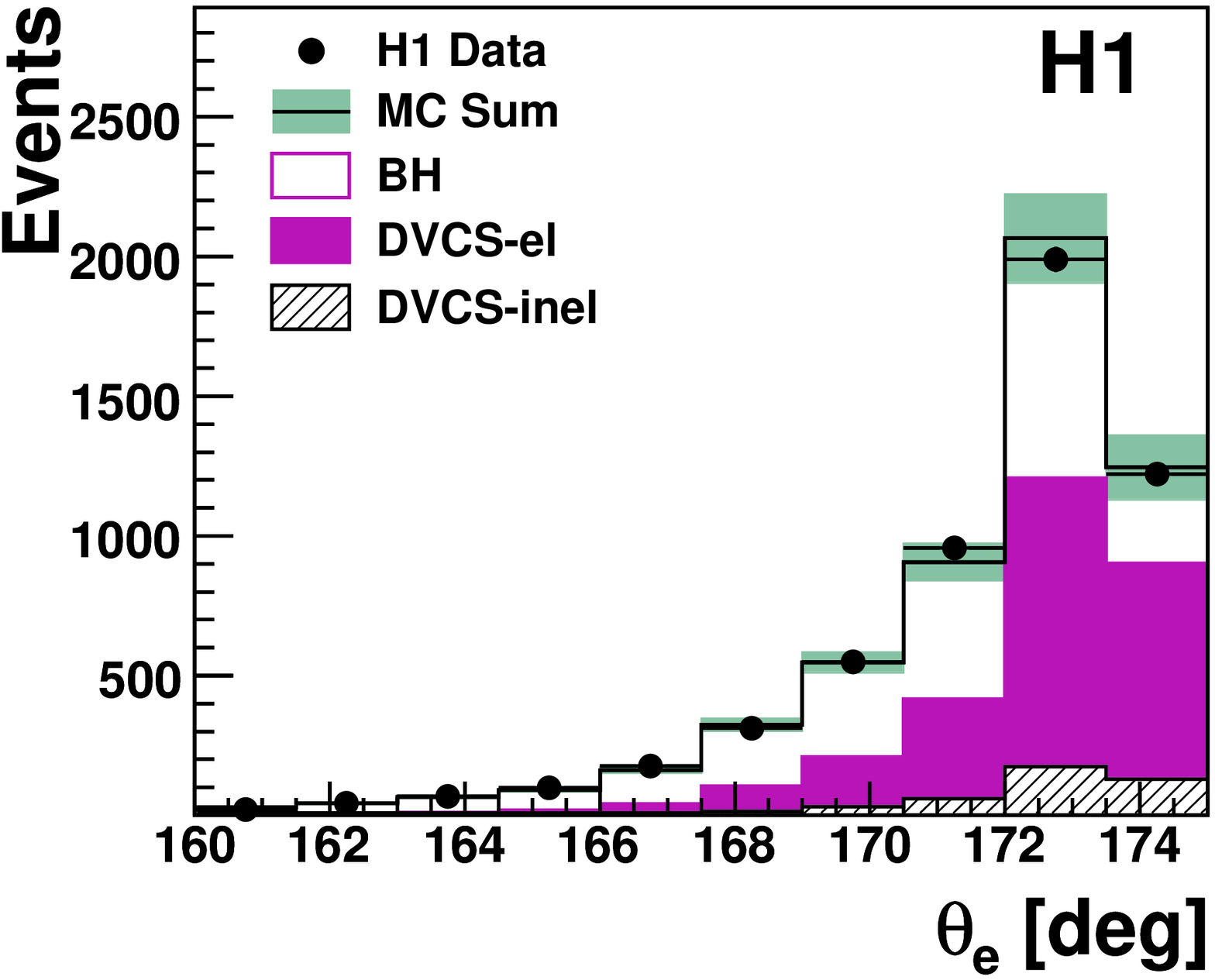}\put(-55,30){{(b)}}\\
 \includegraphics[totalheight=5.7cm]{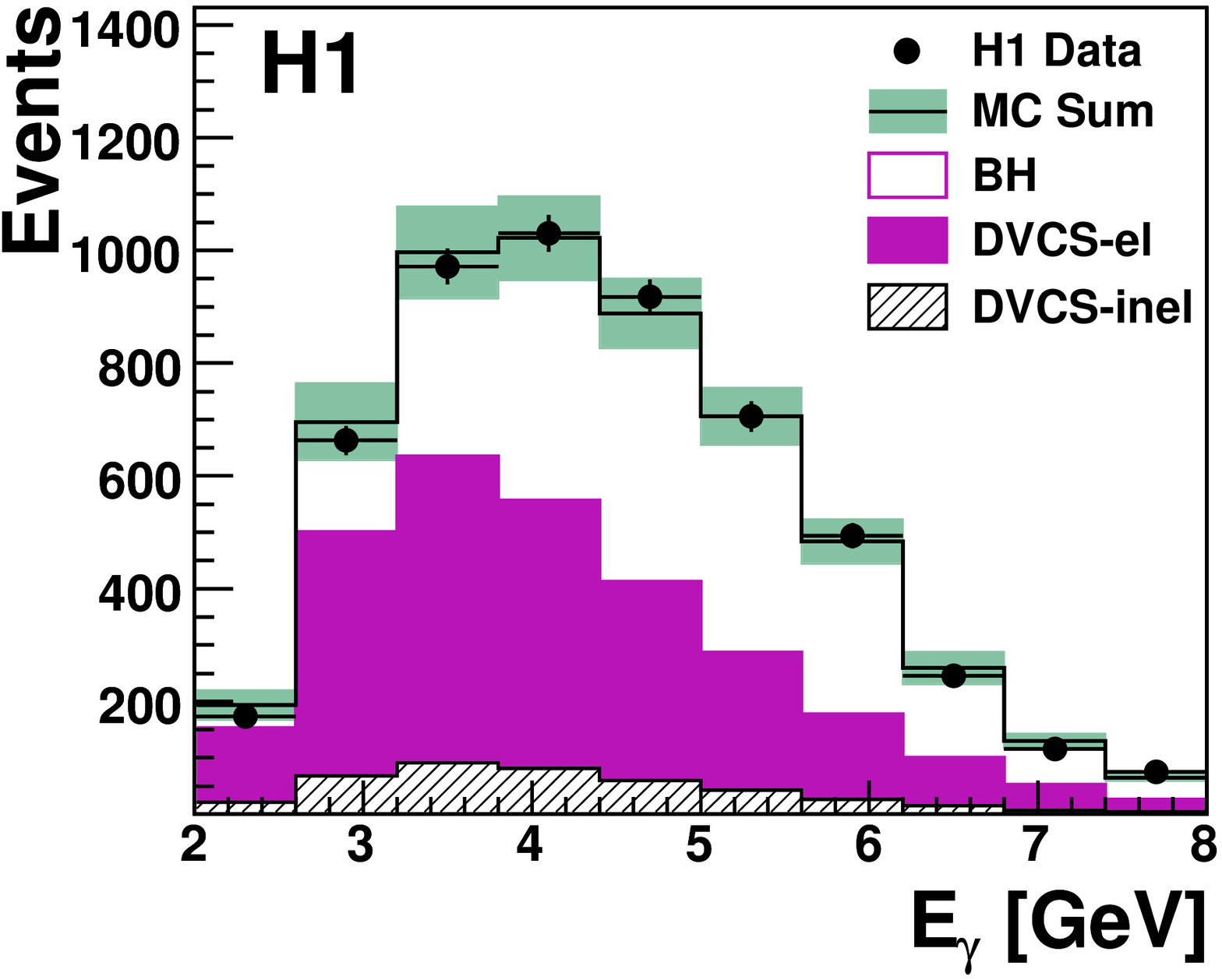}\put(-55,40){{(c)}}
 \includegraphics[totalheight=5.7cm]{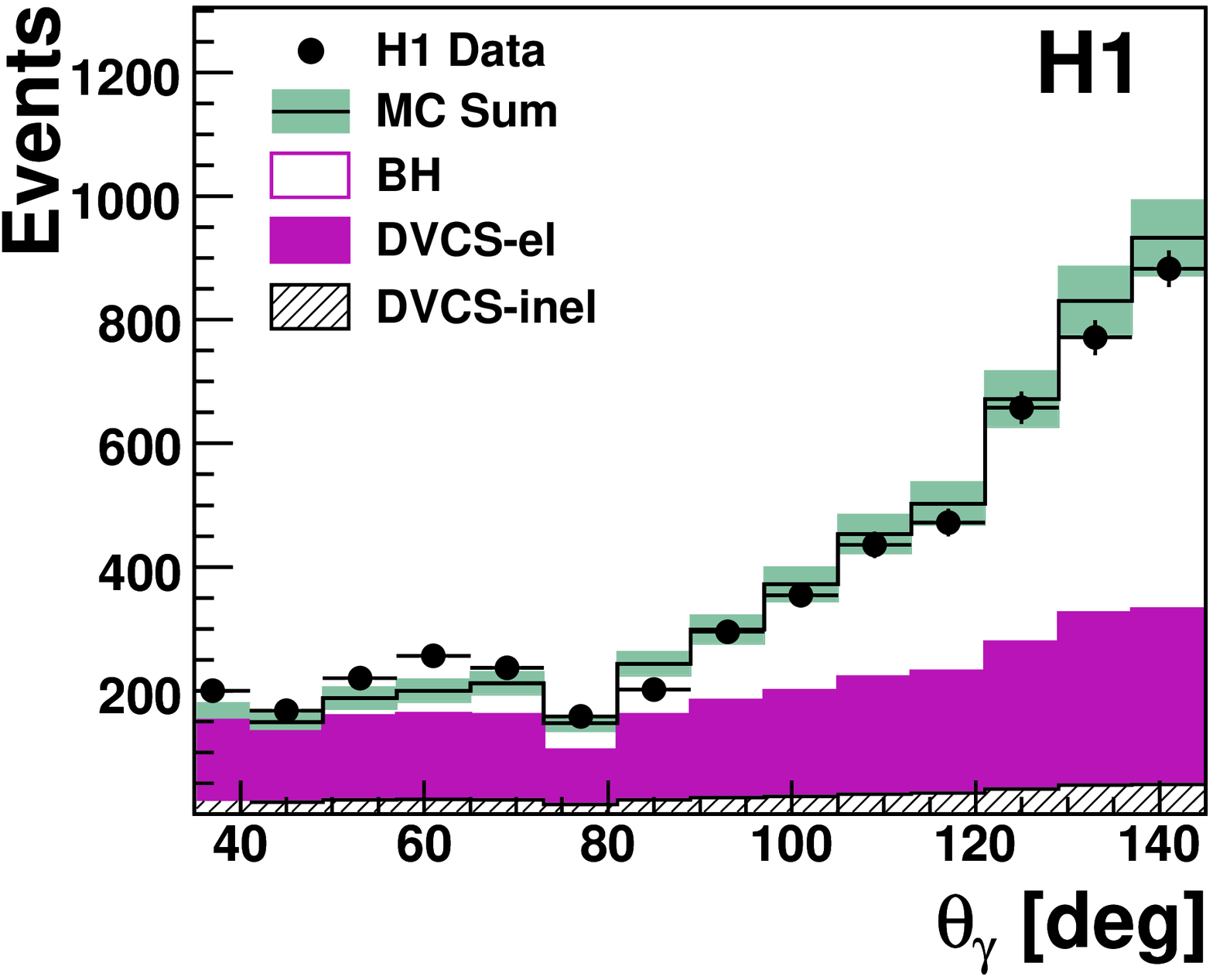}\put(-55,30){{(d)}}\\
 \includegraphics[totalheight=5.7cm]{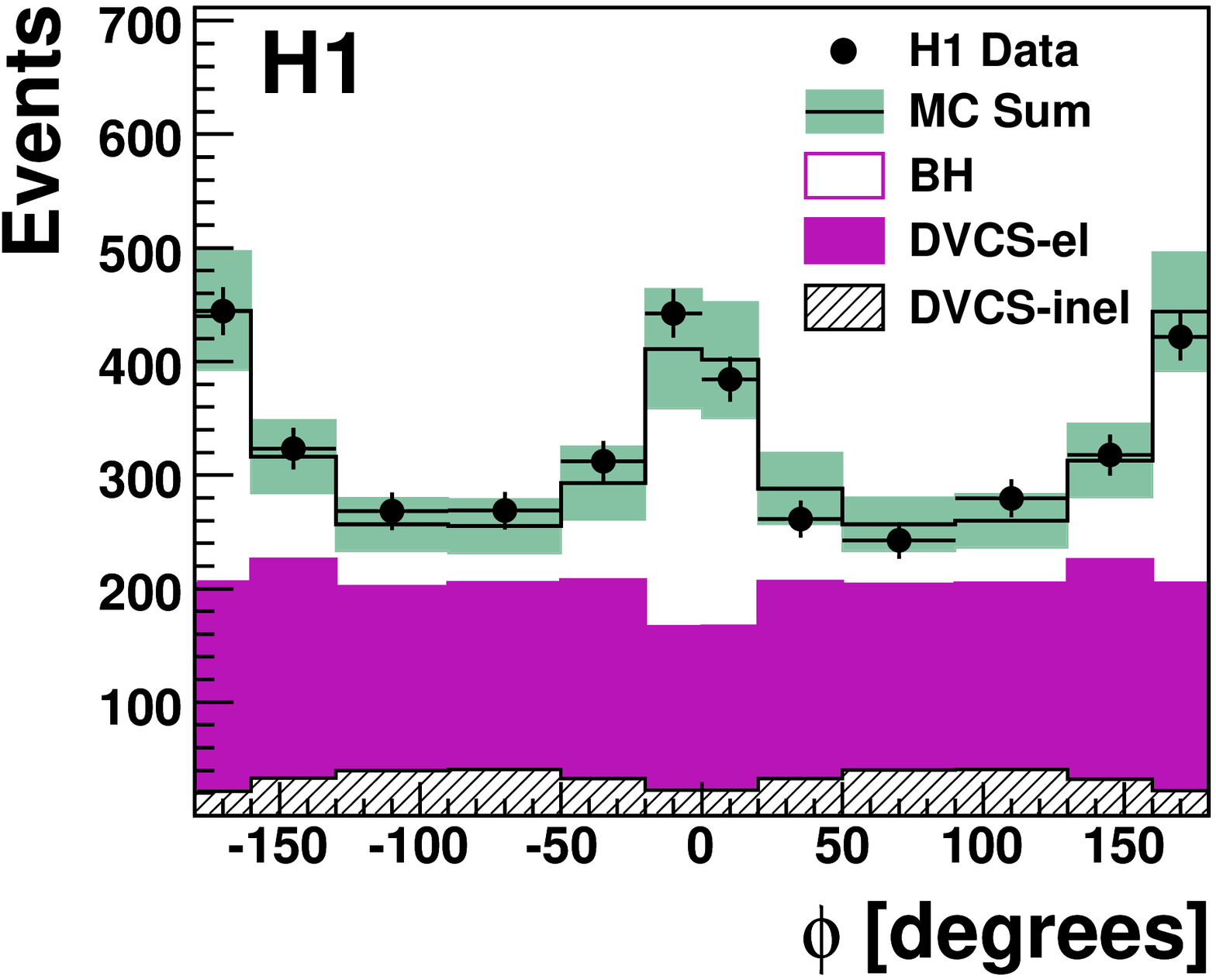}\put(-55,40){{(e)}}
 \includegraphics[totalheight=5.7cm]{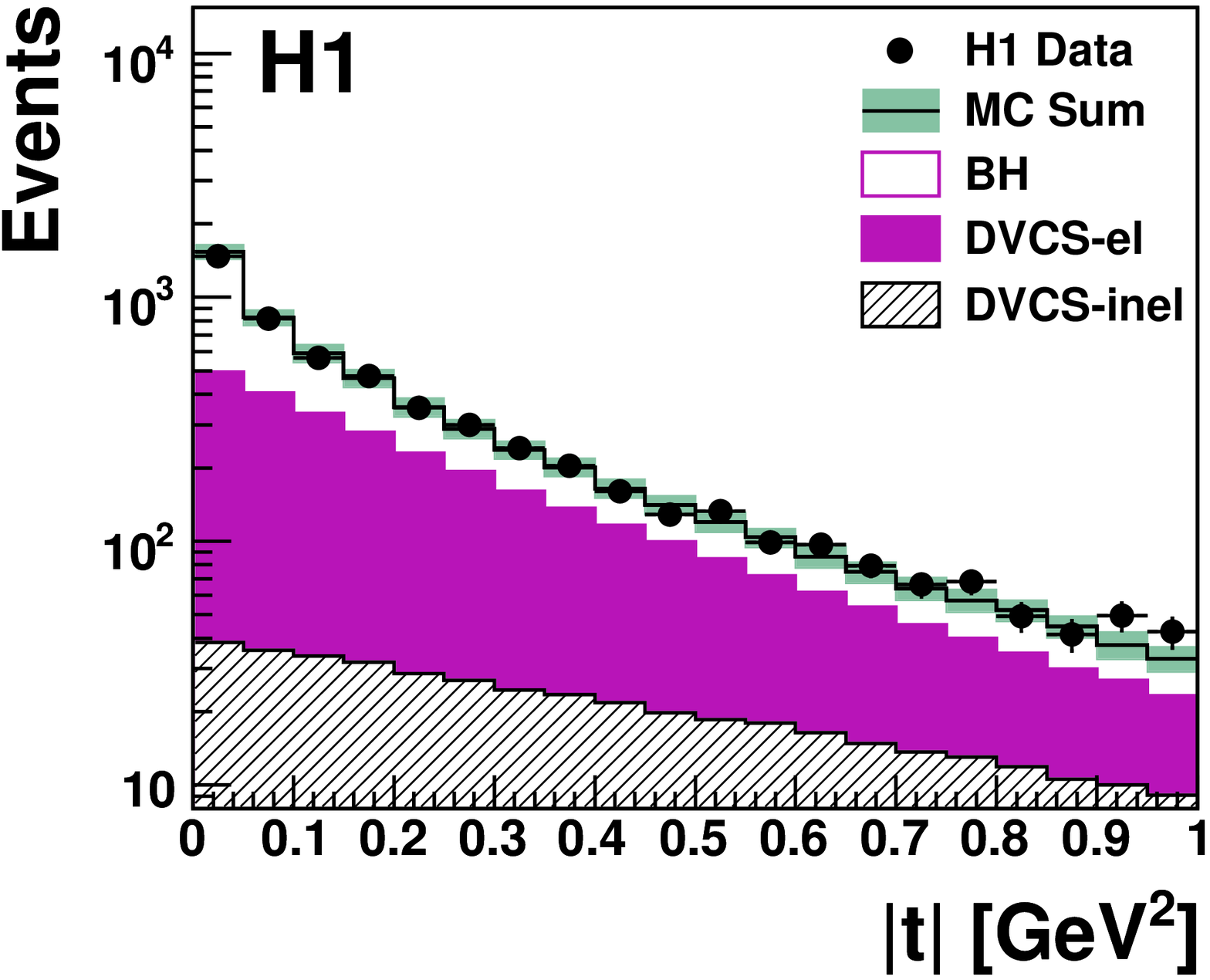}\put(-55,40){{(f)}}
\end{center}
\caption{\label{figcomp} 
Distributions of the energy (a) and polar angle (b) of the scattered
electron,  
the energy (c) and polar angle (d) of the photon, 
the $\phi$ azimuthal angle between the plane of incoming and outgoing lepton and the plane of virtual and real photon~\cite{bel} (e)
and the proton four momentum transfer squared $|t|$ (f).
The data correspond to the full $e^{\pm}p$ sample and are compared to Monte Carlo expectations for 
elastic DVCS, elastic and inelastic BH and inelastic DVCS.
All Monte Carlo simulations are normalised according
to the luminosity of the data.
The open histogram shows the total
prediction and the shaded band its
estimated uncertainty.
}
\end{figure}

The selected event sample contains  $2643$ events in $e^+p$
and $2794$ events in $e^-p$ collisions, respectively.
Distributions of selected kinematic variables are presented in figure~\ref{figcomp} for the full sample from $e^\pm p$ collisions and compared to MC expectation normalised  to the data luminosity.
A good description of the shape and normalisation of the measured distributions is observed.
The analysis sample contains contributions from the elastic DVCS and BH processes, as well as backgrounds from the BH and DVCS processes with proton dissociation, $e  p \rightarrow e  \photon  Y$, where the baryonic system $Y$ of mass $M_Y$ is undetected.
The sum of the latter contributes to $14 \pm 4$\% of the analysis sample, as estimated from MC predictions.
Backgrounds from diffractive 
\om\ and \ph\ production decaying to final states with photons
are estimated to be negligible in the kinematic range of the analysis.
Contamination from processes with low multiplicity $\pi^0$ production was also investigated and found to be negligible.
%

\section{Cross Section and Beam Charge Asymmetry Measurements}
\label{sec:xsections}

The full $e^{\pm}p$ data sample is used to measure the DVCS cross section
integrated over $\phi$. 
The separate $e^+p$ and $e^-p$ data samples are used to measure the beam charge asymmetry as a function of $\phi$.

The DVCS cross section, $\gamma^\ast p \to \gamma p$, is evaluated in each 
bin $i$ at the bin centre values $Q^2_i,W_i,t_i$ using the expression

\begin{equation}
\sigma_{DVCS}(Q^2_i,W_i,t_i)=
\frac
{(N_i^{\rm{obs}}-N_i^{\rm{BH}} - N_i^{\rm{DVCS-inel}})}{N_i^{\rm{DVCS-el}}}\cdot 
\sigma^{\rm{\gamma^\ast p}}_{DVCS-el}(Q^2_i,W_i,t_i) \, ,
\label{eq-ffs}
\end{equation}

\noindent where $N^{\rm{obs}}_i$ is the number of data events observed in 
bin $i$. The other numbers in this equation are calculated using the MC 
simulations described in section~\ref{simul}.
$N^{\rm{BH}}_i$ denotes the number of BH events (elastic and inelastic) reconstructed in bin $i$ and normalised to the data luminosity, $N^{\rm{DVCS-inel}}_i$ the number of inelastic DVCS background events,
$N_i^{\rm{DVCS-el}}$ the number of elastic DVCS events and
$\sigma^{\rm{\gamma^\ast p}}_{DVCS-el}$ is the theoretical $\gamma^\ast p \to \gamma p$ cross section used for the generation of DVCS events. 
The mean value of the acceptance, defined as  the number of DVCS MC events 
reconstructed in a bin divided by the number of events generated in the same bin, is $60$\% over the whole kinematic range, 
for both beam charges. 

The  systematic errors of the measured DVCS cross section
are determined by repeating the analysis after applying to the MC samples appropriate variations for each error source.
The main contribution comes from  the
variation of the $t$-slope parameter set in the elastic DVCS MC by $\pm 6$\%, as constrained by this analysis, and the $4$\% uncertainty of the FMD veto efficiency.
These error sources result in an error of $10$\% on the measured cross section.
The $20$\% uncertainty of the $t$-slope parameter needed to estimate the inelastic DVCS background (see section~\ref{sec:pdiss}) translates into an error on the elastic cross section of $4$\% on average, but reaches $12$\% at high $t$.
The modelling of BH processes by the MC simulation is controlled using the
method detailed in~\cite{dvcsh1} and is attributed an uncertainty of $3$\%.
The uncertainties related to trigger efficiency, photon identification efficiency, 
radiative corrections and
luminosity measurement are each in the range of $1$ to $3$\%.
The total systematic uncertainty of the cross section amounts to 
about $12$\%. A fraction of about $85$\%  of this error is correlated among bins.

For the BCA measurement, the angle $\phi$ is calculated from the reconstructed four-vectors of the electron and of the photon.
MC studies indicate that the resolution of $\phi$ is in the range from $20^\circ$ to $40^\circ$. 
The resolution of $\phi$ is limited mainly by the resolution on the photon energy in the LAr and the resolution on the electron polar angle.
In addition there are large migrations between the true and the reconstructed $|\phi|$ from $0^\circ$ to $180^\circ$, and vice versa.
The asymmetry $A_C(\phi)$ is then determined from the differential $ep \rightarrow ep \gamma$ cross sections $d\sigma^+/d\phi$ and $d\sigma^-/d\phi$ using the formula~(\ref{bcadef}).
The cross sections $d\sigma/d\phi$ are evaluated similarly to $\gamma^\ast p \to \gamma p$ cross section at bin centre values $\phi_i$ using the expression

\begin{equation}
d\sigma/d\phi(\phi_i)=
\frac
{(N_i^{\rm{obs}}-N_i^{\rm{BH-inel}} - N_i^{\rm{DVCS-inel}})}{(N_i^{\rm{DVCS-el}}+N_i^{\rm{BH-el}})}\cdot 
(\sigma^{\rm{ep}}_{DVCS-el}(\phi_i)+\sigma^{\rm{ep}}_{BH-el}(\phi_i)),
\label{eq:xsec_phi}
\end{equation}

\noindent where $N_i^{\rm{BH-el}}$ and $N_i^{\rm{BH-inel}}$ are the numbers of elastic and inelastic MC BH events, respectively, and $\sigma^{\rm{ep}}_{DVCS-el}(\phi_i)+\sigma^{\rm{ep}}_{BH-el}(\phi_i)$ denotes the sum of the theoretical DVCS and BH $ep \rightarrow ep \gamma$ cross sections.
Since a $\cos \phi$ dependence is expected, events with $\phi <0$ and $\phi >0$ are combined, 
in order to increase the statistical significance and to remove effects on the 
asymmetry of any possible $\sin \phi$ contribution from the residual lepton beam polarisation. 
The systematic error on the BCA measurement mainly arises from the part of the LAr photon energy scale uncertainty which is correlated between the $e^+p$ and $e^-p$ samples, estimated to be $0.5$\%.
It leads to sizeable systematic errors on the measured asymmetry for $\phi$ close to $0^\circ$ and $180^\circ$.

In a first step, the interference term between DVCS and BH processes, which is not known a priori, is not included in formula~(\ref{eq:xsec_phi}).
In order to simulate the interference term, an asymmetry of the form $p_1 \cos \phi$ is added to the MC generation and passed through the full detector simulation and analysis chain to account for all acceptance and migration effects from true to reconstructed $\phi$ values.
Similarly to the data, formulae~(\ref{eq:xsec_phi}) and (\ref{bcadef}) are used to determine the reconstructed asymmetry corresponding to these MC events.
To determine the value of $p_1$, a $\chi^2$  minimisation is performed as a function of $p_1$ to adjust the reconstructed asymmetry in the MC to the measured one. 
MC events generated using this $p_1$ value are then used to correct the measured asymmetry for the effect of migrations. Bin by bin correction factors are determined from the difference between the true and the reconstructed asymmetry in the MC.

\section{Results and Interpretations}

\subsection{Cross Sections and $t$-dependence}\label{sec:exp_xsec}

\begin{figure}[!htbp]
\begin{center}
 \includegraphics[totalheight=6cm]{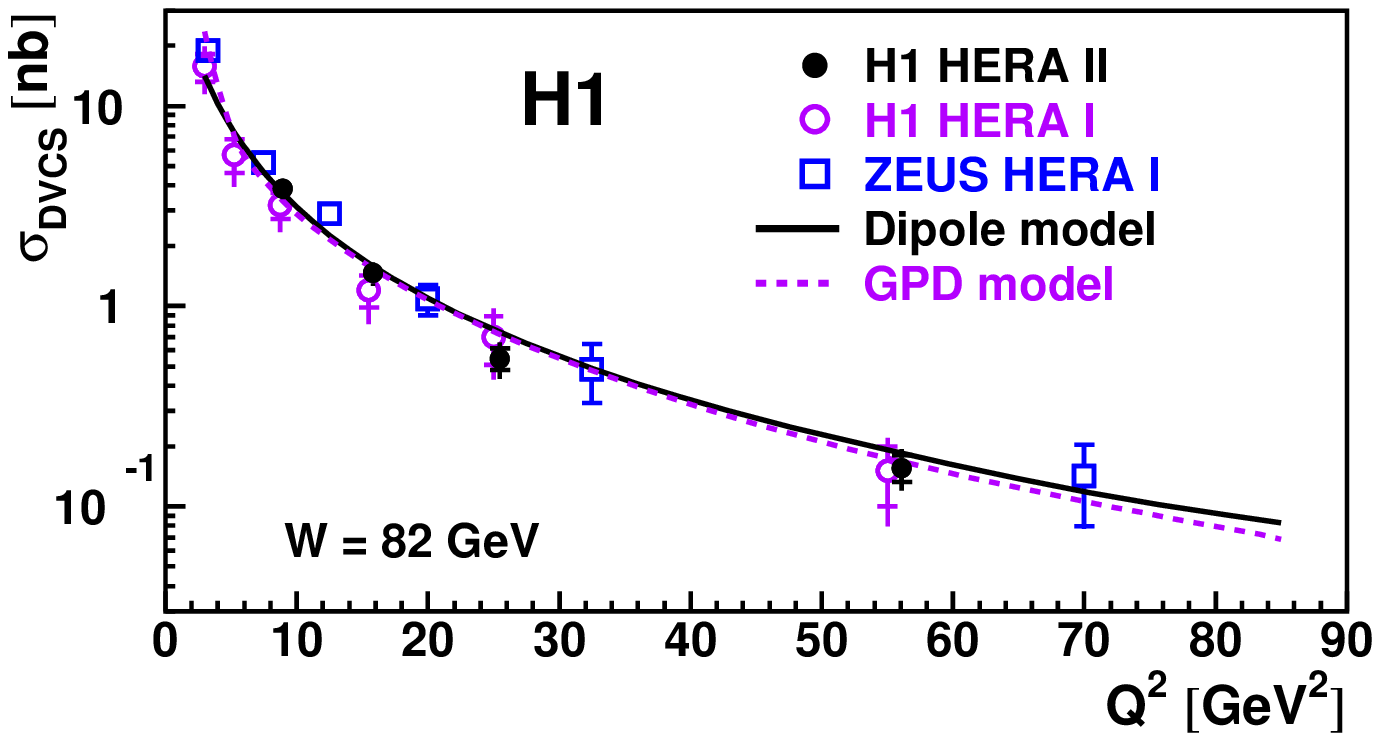}\put(-10,48){{(a)}}\\
 \includegraphics[totalheight=6cm]{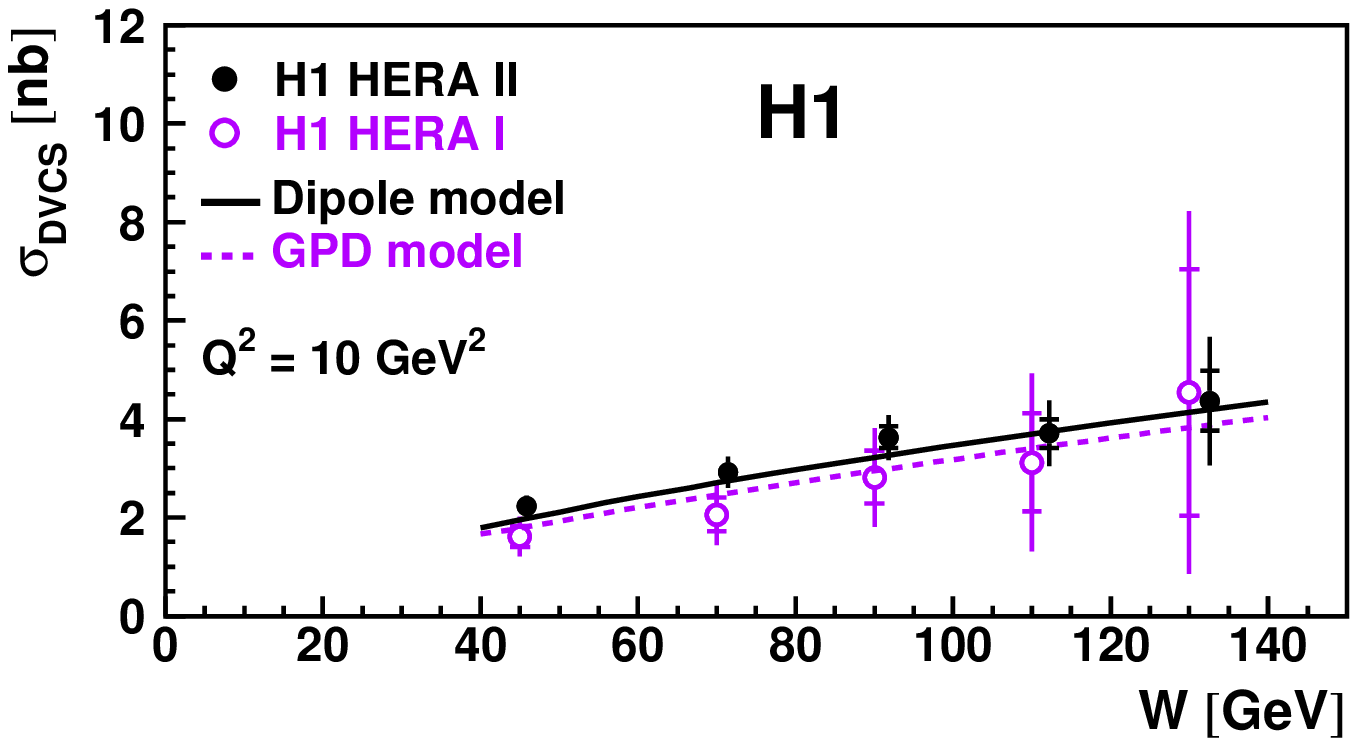}\put(-10,48){{(b)}}
\end{center}
\caption{\label{fig1d} 
The DVCS cross section  $\gamma^\ast p \rightarrow \gamma p$ as a function of
$Q^2$ at  $W=82$~GeV (a) and as a function of
$W$ at $Q^2=10$~GeV$^2$ (b).
The results from the previous H1~\cite{dvcsh1a} and ZEUS~\cite{dvcszeusb} publications 
based on HERA~I data are also displayed.
ZEUS measurements are propagated from $W=104$~GeV to $82$~GeV using a $W$ dependence $W^{0.52}$.
The inner error bars represent the statistical errors, 
the outer error bars the statistical and systematic errors added in quadrature.
The dashed line represents the prediction of the GPD model~\protect{\cite{muller}} and the solid line the prediction of the dipole model~\protect{\cite{gregory}}.
}
\end{figure}

The measured DVCS cross sections as a function of $W$ for $|t| < 1$~GeV$^2$ and at $Q^2 = 10 $~GeV$^2$ as well as the $Q^2$ dependence at $W = 82$~GeV are displayed in figure~\ref{fig1d} and given in table~\ref{sig1d}. 
They agree within errors with the previous measurements~\cite{dvcsh1,dvcsh1a,dvcszeus,dvcszeusb}. 
The data agree also with models based on GPDs~\cite{muller}  or the dipole approach~\cite{gregory}. 
DVCS cross sections for $e^{+}p$ and $e^{-}p$ data 
are also found in good agreement with each other.
As already discussed in~\cite{dvcsh1},
the steep rise of the cross section with $W$ 
is an indication of the presence of a hard underlying process.

\begin{figure}[!htbp]
\begin{center}
 \includegraphics[totalheight=6cm]{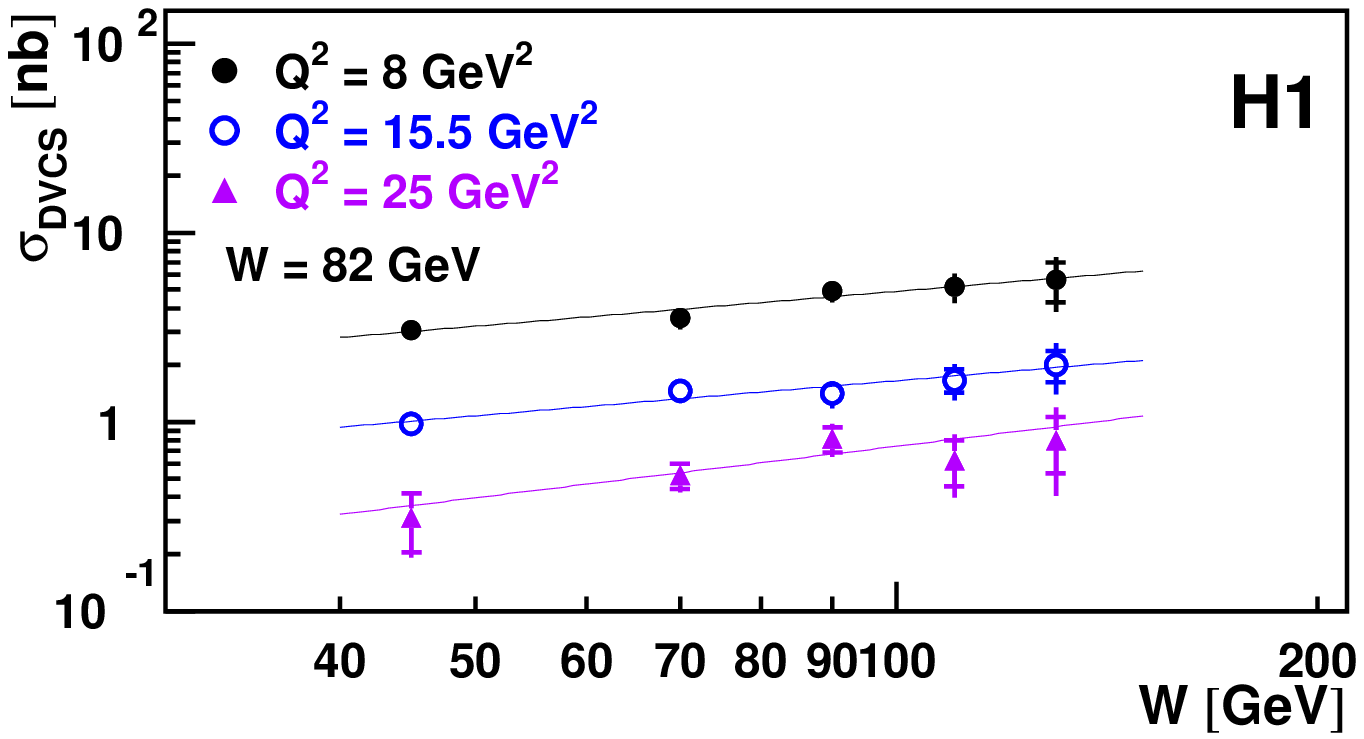}\put(-10,15){{(a)}}\\
 \includegraphics[totalheight=6cm]{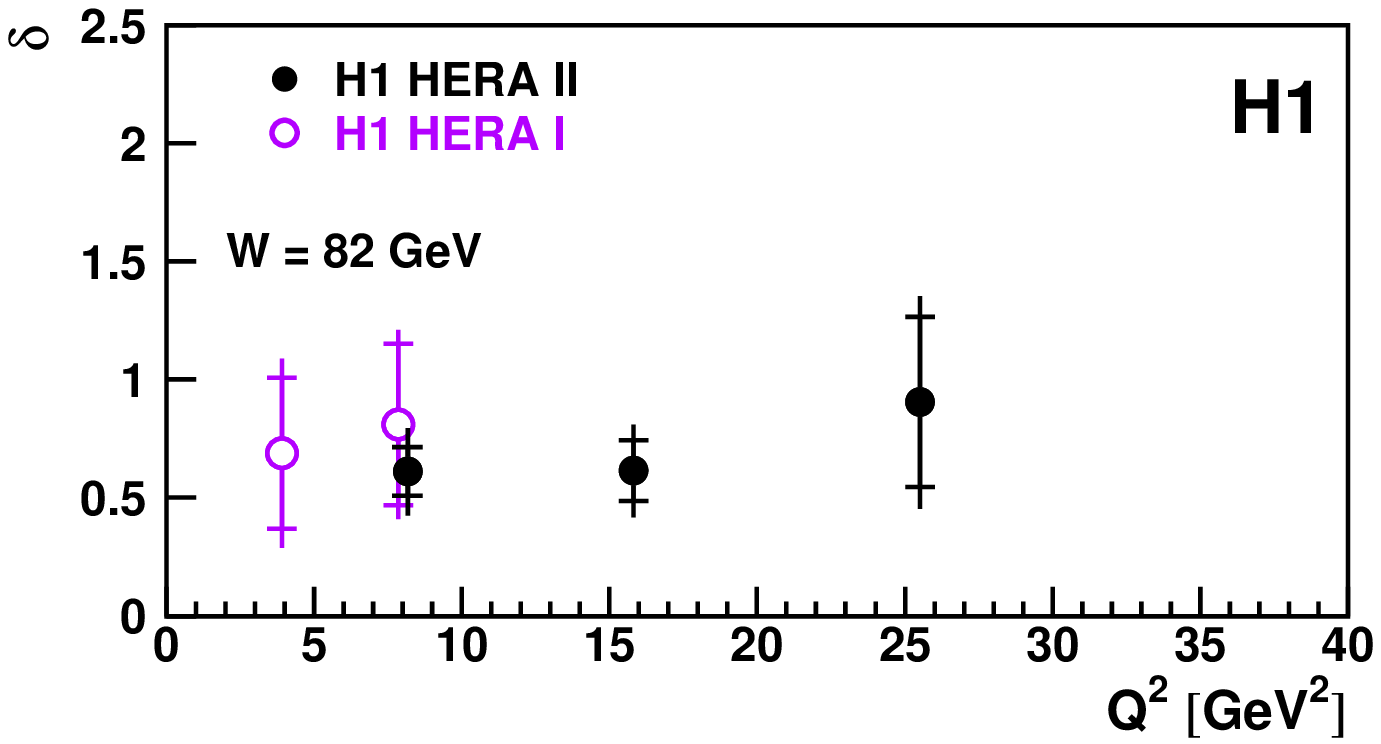}\put(-10,15){{(b)}}
\end{center}
\caption{\label{fig2d} 
The DVCS cross section $\gamma^\ast p \rightarrow \gamma p$ as a function of
$W$ at three values of $Q^2$ (a). The solid lines represent the results
of fits of the form $W^\delta$. The fitted values of $\delta(Q^2)$
are shown in (b) together with the values obtained using HERA I data~\protect{\cite{dvcsh1a}}. 
The inner error bars represent the statistical errors, 
the outer error bars the statistical and systematic errors added in quadrature.
}
\end{figure}

The $W$ dependence of the cross section for three separate bins of $Q^2$ is shown in figure~\ref{fig2d}(a) and given in table~\ref{sig2d}.
A fit of the function $W^\delta$ is performed in each $Q^2$ bin. Figure~\ref{fig2d}(b) shows the obtained $\delta$ values. 
It is observed that $\delta$ is independent of $Q^2$ within the errors.
The average value\footnote{Here and in all other places where results are given the first error is statistical and the second systematic.} $\delta = 0.63 \, \pm \, 0.08 \, \pm \, 0.14$ is in agreement with the previous measurement~\cite{dvcsh1}, as well as with the value of $\delta = 0.52 \pm 0.09$~(stat.) measured by the ZEUS Collaboration at a lower $Q^2$ of $3.2$~GeV$^2$~\cite{dvcszeusb}.

\begin{figure}[!htbp]
\begin{center}
 \includegraphics[width=9.5cm]{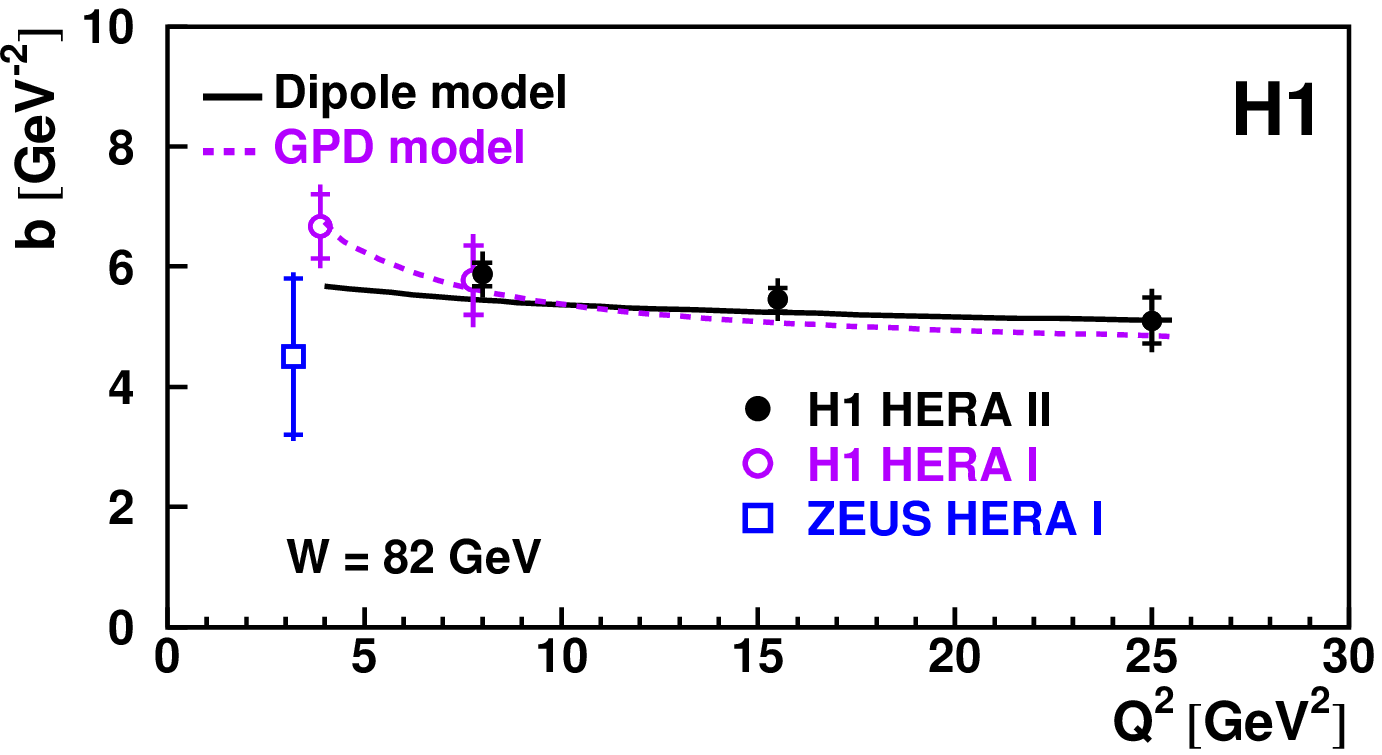}\put(-10,15){{(a)}}\\
 \includegraphics[width=9.5cm]{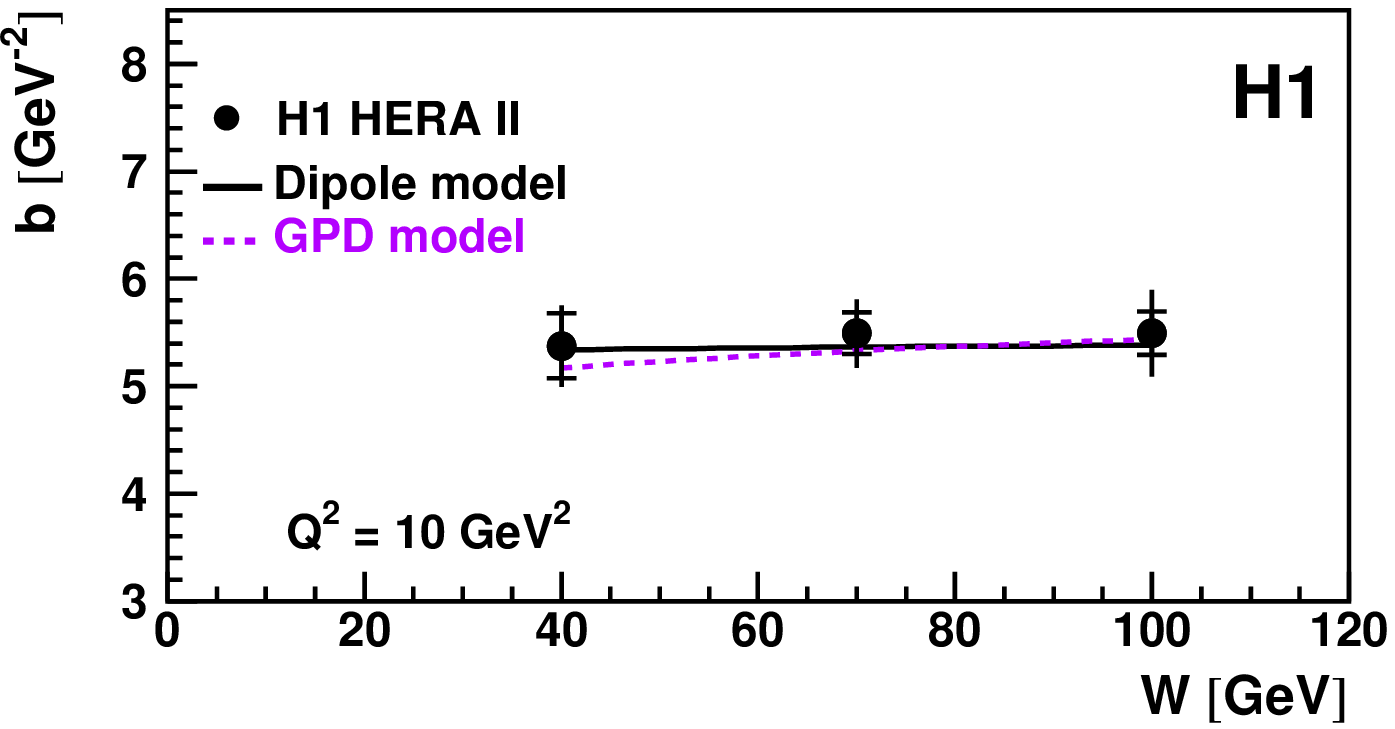}\put(-10,15){{(b)}}
\end{center}
\vspace*{-0.5cm}
\caption{\label{figb} 
The fitted $t$-slope parameters $b(Q^2)$ are shown in (a) together with the
$t$-slope parameters from the previous H1~\cite{dvcsh1a} and ZEUS~\cite{dvcszeusb} publications 
based on HERA I data.
In (b) the fitted $t$-slope parameters  $b(W)$ are shown. 
The inner error bars represent the statistical errors
and the outer error bars the statistical and systematic errors added in quadrature.
The dashed line represents the prediction of the GPD model~\protect{\cite{muller}} and the solid line the prediction of the dipole model~\protect{\cite{gregory}}.
}
\end{figure}

Differential cross sections are measured as a function of $t$ for three values of $Q^2$ and $W$
and presented in table~\ref{sigtq}.
Fits of the form $d\sigma/d|t| \sim e^{-b|t|}$, which describe the data well~\cite{dvcsh1}, are performed taking into account the statistical and correlated systematic errors.
The derived $t$-slope parameters $b(Q^2)$ and $b(W)$ are displayed in figures~\ref{figb}(a) and (b), respectively.
They confirm the result obtained in a previous analysis \cite{dvcsh1} and no  
significant variation of $b$ with $W$ is observed.
Experimental results are compared with calculations from GPD and dipole models~\cite{muller,gregory}.
A good agreement is obtained for both $W$ and $Q^2$ dependences of the $t$-slopes.
It should be noted that in the GPD model previous data of~\cite{dvcsh1,dvcsh1a} are used to derive the $Q^2$ and $W$ dependences of $b$, while no DVCS data enter in the determination of parameters of the dipole model.
If $b$ is parametrised as \mbox{$b=b_0+2 \alpha' \ln \frac{1}{x}$}, with $x=Q^2/W^2$, the obtained $\alpha'$ value is compatible with $0$ and 
an upper limit on $\alpha'$ of $0.20$ GeV$^{-2}$ at $95$\% confidence level (CL) is derived.
This value is compatible with results obtained for $\jpsi$ exclusive electroproduction~\cite{jpsih1,jpsizeus}, 
for which the measured $\alpha'$ is below $0.17$ GeV$^{-2}$ at $95$\% CL.
An increase of the slope with decreasing $x$ (shrinkage) is therefore not observed. 
Such a behaviour is expected for hard processes and confirms that perturbative QCD can be used to describe DVCS processes.

Using the complete analysis sample, the value of $b$ is 
found to be $5.41 \, \pm \, 0.14 \, \pm \, 0.31$~GeV$^{-2}$ at $Q^2 = 10\; \mbox{GeV}^2$.
This corresponds to a total uncertainty of $6$\% on the (elastic) $t$-slope 
measurement for the full data sample.
As in \cite{dvcsh1}, this $t$-slope  value can be converted to an average impact parameter of 
$\sqrt{<r_T^2>} = 0.64 \pm 0.02$~fm. 
It corresponds to the transverse extension
of the parton density, dominated by sea quarks and gluons for an average value \mbox{$x =1.2 \cdot 10^{-3}$}, 
in the plane perpendicular to the direction of motion of the proton. 
At larger values of $x$ ($x>0.1$), a smaller value of $\sqrt{<r_T^2>}$, dominated by the contribution of valence quarks, is estimated~\cite{diehl}.

\subsection{Inelastic DVCS $t$-dependence}\label{sec:pdiss}

\begin{figure}[!htbp]
\begin{center}
 \includegraphics[width=9.5cm]{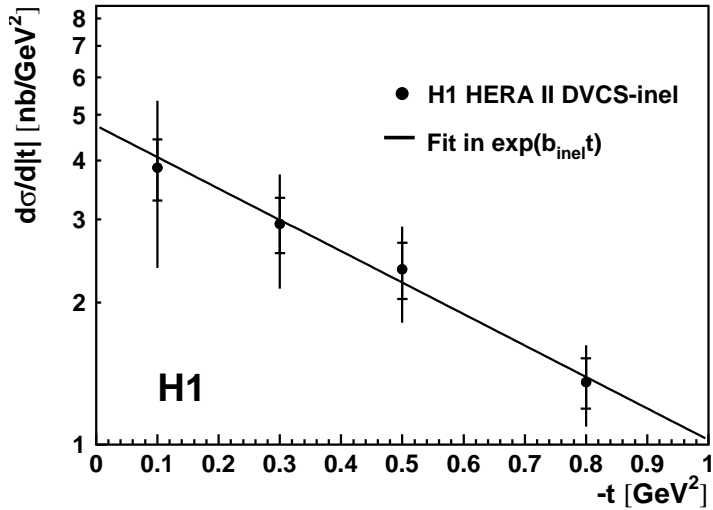}\\
\end{center}
\caption{\label{fig_pdiss} 
The inelastic DVCS cross section  differential in
$t$ at  $W=82$~GeV and $Q^2=10$~GeV$^2$ and for events with $1.4 \lesssim M_Y \lesssim 10$~GeV.
The inner error bars represent the statistical errors, 
the outer error bars the statistical and systematic errors added in quadrature.
}
\end{figure}

The increased statistical precision compared to previous analyses allows a first measurement of the $t$-slope of the inelastic DVCS process.
A sample of events with a signal in the FMD is selected.
It corresponds to events with the mass of the proton dissociation system $M_Y$ in the range $1.4$ to $10$~GeV, as derived from MC studies.
The contribution of inelastic DVCS events is extracted by subtracting the BH (elastic and inelastic) and elastic DVCS contributions, as estimated from the respective MC expectations.
The measured differential cross section as a function of $t$ is presented in figure~\ref{fig_pdiss}.
A fit of the form  $d\sigma/d|t| \sim e^{-b_{inel} |t|}$ yields $b_{inel}=1.53 \pm 0.26 \pm 0.44$~GeV$^{-2}$.
In the present event sample, no indication of a dependence of $b_{inel}$ with $Q^2$ or $W$ is observed.
The obtained value for $b_{inel}$ is compatible with previous determinations for inelastic
exclusive production of $\rho$,
 $\phi$~\cite{Adloff:1997jd} and $\jpsi$~\cite{jpsizeus}.
%

\subsection{Beam Charge Asymmetry}

\begin{figure}[!htbp] 
  \begin{center}
    \includegraphics[width=10cm]{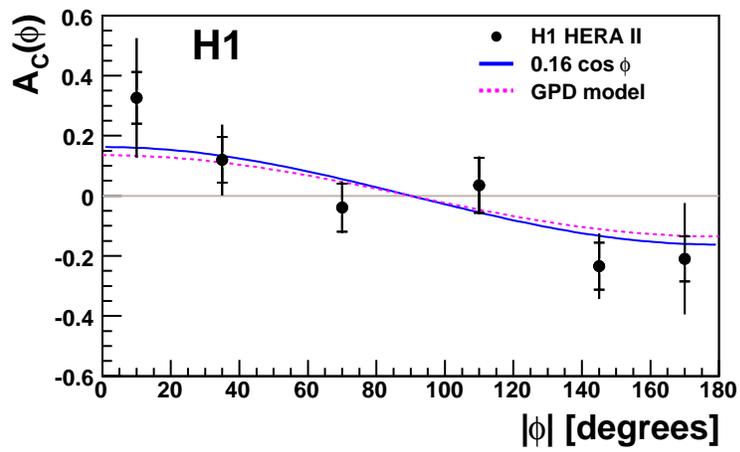}
  \end{center}
  \caption{
Beam charge asymmetry as a function of the angle $\phi$ as defined in~\cite{bel}, integrated over the kinematic range of the analysis.
The inner error bars represent the statistical
errors, the outer error bars the statistical and systematic errors added in
quadrature.
The function  $0.16 \cos\phi$ is also shown (solid line),
together with the GPD model prediction (dashed line).
}
\label{fig3}  
\end{figure}

The contributions of elastic DVCS and BH processes to the analysis sample are of similar size, as can be observed in figure~\ref{figcomp}. This is a favourable situation for the beam charge asymmetry measurement, with a maximum sensitivity for the interference term.
The measured BCA integrated over the kinematic range of the analysis and  corrected for detector effects, as detailed in section~\ref{sec:xsections}, is presented in figure~\ref{fig3} and table~\ref{tabbca}.
Bins in $\phi$ with a size of the order of the experimental resolution on $\phi$ are used.

The $\chi^2$ minimisation procedure leads to a $p_1$ value of \mbox{$p_1 = 0.16 \pm 0.04 \pm 0.06$}.
The resulting function $0.16 \cos \phi$ is displayed in figure~\ref{fig3} and is seen to agree with the prediction of the GPD model for the first $\cos \phi$ harmonic~\cite{muller}.
The measured asymmetry is in good agreement with the model prediction within experimental errors. 

As detailed in section~\ref{sec:theo},
from the measured BCA and the $p_1$ value determined above,
together with the DVCS cross section, the ratio $\rho$  
 of the real to imaginary parts of the DVCS amplitude
can be calculated as $\rho = 0.20 \pm 0.05 \pm 0.08$.
This is the first measurement of this ratio. 
The dispersion relation of equation~(\ref{dispersion}) and our measurement of $\delta(Q^2)$ on the other hand leads to $\rho = 0.25 \pm 0.03 \pm 0.05$,
in good agreement with the direct determination.
While in the low $x$ domain of the present measurement, the real part of the DVCS amplitude is positive, in contrast, at larger $x$ ($x \sim 0.1$) and lower $Q^2$, a smaller and negative real part was measured\footnote{The convention used in~\cite{hermes} for the definition of the $\phi$ angle is different from the one of~\cite{bel} adopted in the present paper.} by the HERMES Collaboration~\cite{hermes}.

\section{Conclusion}

The elastic DVCS cross section $\gamma^\ast p \rightarrow \gamma p$
has been measured with the H1 detector at HERA.
The~~measurement~~is~~performed~~in the 
kinematic~~range~~$6.5\;<\;Q^2\;<\;80$~GeV$^2$, \mbox{$30~< W<~140$~GeV} and $|t| <$~1~GeV$^2$. 
The analysis uses  $e^{+}p$ and  $e^{-}p$ data recorded from $2004$ to $2007$, corresponding to a total integrated luminosity of $306$~pb$^{-1}$, almost equally shared between both beam charges.
The $W$ dependence of the DVCS cross section is well described by a function
$W^{\delta}$.
No significant variation of the exponent $\delta$ as a function of $Q^2$ is observed.
For the total sample a value $\delta = 0.63 \, \pm \, 0.08 \, \pm \, 0.14$ is determined.
The steep rise of the cross section with $W$ indicates a hard underlying process.
The $t$-dependence of the cross section is well described by the form $e^{-b|t|}$
with an average slope of 
$b = 5.41 \, \pm \, 0.14 \, \pm \, 0.31$~GeV$^{-2}$. 
The $t$-slopes are determined differentially in $Q^2$ and $W$ and are
compatible with previous observations.
The $t$-slope is also measured for the inelastic DVCS.
The measured elastic DVCS cross section is compared to the predictions of two different models based on GPDs or on a dipole approach, respectively. Both approaches describe the data well.
The use of $e^{+}p$ and  $e^{-}p$ collision data allows 
the measurement of the beam charge asymmetry of the interference between the BH and DVCS processes, for the first time at a collider.
The ratio $\rho$ of the real to imaginary part of the DVCS amplitude is then derived, directly from the
measurements of the BCA and of the DVCS cross section to be $\rho = 0.20 \pm 0.05  \pm 0.08$.
This ratio can also be calculated from a dispersion relation using only the DVCS energy dependence, leading to $\rho = 0.25 \pm 0.03  \pm 0.05$.
Both results are in good agreement.
The GPD model considered here~\cite{muller}
 correctly describes the measured BCA as well as $\rho$.
The measurements presented here show that a combined analysis of DVCS observables, including cross section and charge asymmetry,
allows the extraction of the real part of the DVCS amplitude and subsequently a
novel understanding of the correlations of parton momenta in the proton.

\section*{Acknowledgements}

We are grateful to the HERA machine group whose outstanding
efforts have made this experiment possible. 
We thank the engineers and technicians for their work in constructing 
and maintaining the H1 detector, our funding agencies for financial 
support, the DESY technical staff for continual assistance and the 
DESY directorate for the hospitality which they extend to the non DESY 
members of the collaboration.
We would like to thank Dieter Mueller, Kresimir Kumeri\v{c}ki and
Gregory Soyez for helpful discussions and for providing theory predictions.



\vfill
\newpage

\begin{table}[htbp]
\centering
\begin{tabular}{|c|lll|l|c|lll|}
\cline{1-4} \cline{6-9} 
    & 
   \multicolumn{3}{c|}{} &
\hspace{0.5 cm} &   & 
  \multicolumn{3}{c|}{} \\[-10pt]
  $Q^2$ $\left[{\rm GeV}^2 \right]$  & 
   \multicolumn{3}{c|}{$\sigma_{DVCS}$ $\left[{\rm nb}\right]$} &
\hspace{0.5 cm} & $W$ $\left[{\rm GeV}\right]$  & 
  \multicolumn{3}{c|}{$\sigma_{DVCS}$ $\left[{\rm nb}\right]$} \\[1.5pt]
\cline{1-4} \cline{6-9}
    & 
   \multicolumn{3}{c|}{} &
\hspace{0.5 cm} &   & 
  \multicolumn{3}{c|}{} \\[-12pt]
 $8.75$  & $3.87$  &$\pm$ $0.15 $&$\pm$ $0.41 $ & &  $ 45$ & $2.23$  &$\pm$ $0.11$&$\pm$ $0.19 $ \\
 $15.5$  & $1.46$  &$\pm$ $0.07 $&$\pm$ $0.18 $ & &  $ 70$ & $2.92$  &$\pm$ $0.16$&$\pm$ $0.27 $ \\
 $25$    & $0.55$  &$\pm$ $0.07$ &$\pm$ $0.08 $  & & $ 90$ & $3.63$  &$\pm$ $0.22$&$\pm$ $0.40 $ \\
 $55$    & $0.16$  &$\pm$ $0.02$ &$\pm$ $0.03 $  & & $110$ & $3.71$  &$\pm$ $0.29$&$\pm$ $0.61 $ \\
         &         &           &                 & & $130$ & $4.37$  &$\pm$ $0.60$&$\pm$ $1.16 $ \\[1.5pt]
\cline{1-4} \cline{6-9}
\end{tabular}
\caption{ 
 The DVCS cross section $\gamma^\ast p \rightarrow \gamma p$, $\sigma_{DVCS}$,
 as a function of $Q^2$ for 
 $W=82\,{\rm GeV}$ and as a function of $W$ for $Q^2=10\,{\rm GeV}^2$, both 
 for $ |t| < 1\,{\rm GeV}^2$.
 The first errors are statistical, the second systematic.}
\label{sig1d}
\end{table}

\begin{table}[htbp]
\centering
\begin{tabular}{|c|lcc|lcc|lcc|}
 \cline{2-10}
 \multicolumn{1}{c|}{~} &\multicolumn{9}{c|}{~} \\ [-10pt]
 \multicolumn{1}{l|}{} & \multicolumn{9}{c|}{$\sigma_{DVCS}
    \; \; \left[{\rm nb}\right]$} \\ [3.0pt]
 \hline
 \multicolumn{1}{|c|}{~} &\multicolumn{3}{|c|}{~}  
   &\multicolumn{3}{c|}{~}  
   &\multicolumn{3}{c|}{~} \\ [-10pt]
 $W$ $\left[{\rm GeV}\right] $  &
  \multicolumn{3}{c|}{$Q^2 = 8\,{\rm GeV}^2$} &  
  \multicolumn{3}{c|}{$Q^2 = 15.5\,{\rm GeV}^2$} & 
  \multicolumn{3}{c|}{$Q^2 = 25\,{\rm GeV}^2$} \\ [3.0pt]
 \hline 
      $45$&       $3.06$&    $\pm$   $0.18$&  $\pm$     $0.25$&    $0.98$&   $\pm$    $0.07$&  $\pm$	$0.08$&	    $0.31$&   $\pm$    $0.11$&  $\pm$  $0.05$ \\
      $70$&       $3.54$&    $\pm$   $0.29$&  $\pm$     $0.34$&    $1.46$&   $\pm$    $0.12$&  $\pm$	$0.12$&	    $0.52$&   $\pm$    $0.08$&  $\pm$  $0.06$ \\
      $90$&       $4.93$&    $\pm$   $0.39$&  $\pm$     $0.52$&    $1.41$&   $\pm$    $0.16$&  $\pm$	$0.17$&	    $0.81$&   $\pm$    $0.13$&  $\pm$  $0.09$ \\
     $110$&       $5.16$&    $\pm$   $0.51$&  $\pm$     $0.74$&    $1.66$&   $\pm$    $0.23$&  $\pm$	$0.28$&	    $0.63$&   $\pm$    $0.17$&  $\pm$  $0.15$ \\
     $130$&       $5.62$&    $\pm$   $1.34$&  $\pm$     $1.19$&    $2.00$&   $\pm$    $0.37$&  $\pm$	$0.47$&	    $0.80$&   $\pm$    $0.26$&  $\pm$  $0.29$ \\
 \hline 
 \hline 
  $\delta$  & $0.61$ & $\pm$ $0.10$ & $\pm$ $0.15$ & $0.61$ & $\pm$ $0.13$ & $\pm$ $0.15$ & $0.90$ & $\pm$  $0.36$ &  $\pm$ $0.27$  \\
 \hline 
\end{tabular}
\caption{The DVCS cross section $\gamma^\ast p \rightarrow \gamma p$, $\sigma_{DVCS}$,
as a function of $W$ for  three $Q^2$ values and for $ |t| < 1\,{\rm GeV}^2$.
The values of $\delta(Q^2)$
obtained from fits of the form $W^\delta$ are given.
The first errors are statistical, the second systematic.}
\label{sig2d}
\end{table}

\begin{table}[htbp]
\centering
\begin{tabular}{|c|lcc|lcc|lcc|}
   \cline{2-10}
  \multicolumn{1}{c|}{~} &\multicolumn{9}{c|}{~} \\ [-10pt]
  \multicolumn{1}{l|}{} & \multicolumn{9}{c|}{$d\sigma_{DVCS}/d|t|
     \; \; \left[{\rm nb/GeV}^2\right]$} \\ [3.0pt]
  \cline{2-10}
   \multicolumn{1}{c|}{~} &\multicolumn{9}{|c|}{~} \\ [-11pt]
    \multicolumn{1}{c|}{~} &
   \multicolumn{9}{c|}{$W=82\,{\rm GeV}$} \\ [1.0pt]
\hline
  \multicolumn{1}{|c|}{~} &\multicolumn{3}{|c|}{~}
    &\multicolumn{3}{c|}{~}
    &\multicolumn{3}{c|}{~} \\ [-12pt]
  $|t|$ $\left[{\rm GeV}^2\right] $ &
  \multicolumn{3}{c|}{$Q^2 = 8\,{\rm GeV}^2$} &  
  \multicolumn{3}{c|}{$Q^2 = 15.5\,{\rm GeV}^2$} & 
  \multicolumn{3}{c|}{$Q^2 = 25\,{\rm GeV}^2$} \\ [3.0pt]
 \hline 
 $0.10$&   $13.3$&  $\pm$ $0.80$&  $\pm$   $1.73$&	 $4.33$&    $\pm$  $0.35$&   $\pm$ $0.65$&    $1.68$& $\pm$  $0.31$&   $\pm$ $0.42$ \\
 $0.30$&   $4.82$&  $\pm$ $0.32$&  $\pm$   $0.50$&	 $1.24$&    $\pm$  $0.13$&   $\pm$ $0.16$&    $0.49$& $\pm$  $0.10$&   $\pm$ $0.08$ \\
 $0.50$&   $1.26$&  $\pm$ $0.14$&  $\pm$   $0.18$&	 $0.45$&    $\pm$  $0.06$&   $\pm$ $0.05$&    $0.18$& $\pm$  $0.04$&   $\pm$ $0.03$ \\
 $0.80$&   $0.21$&  $\pm$ $0.03$&  $\pm$   $0.04$&	 $0.10$&    $\pm$  $0.01$&   $\pm$ $0.02$&    $0.05$& $\pm$  $0.01$&   $\pm$ $0.01$ \\
 \hline 
 \hline 
  $b$ [${\rm GeV}^{-2}$]  & $5.87$ & $\pm$ $0.20$ & $\pm$ $0.32$ & $5.45$ & $\pm$ $0.20$ & $\pm$ $0.29$ & $5.10$ & $\pm$ $0.38$ & $\pm$ $0.37$  \\
 \hline 
\multicolumn{10}{c}{}\\
 \cline{2-10}
   \multicolumn{1}{c|}{~} &\multicolumn{9}{|c|}{~} \\ [-11pt]
    \multicolumn{1}{c|}{~} &
   \multicolumn{9}{c|}{$Q^2=10\,{\rm GeV}^2$} \\ [1.0pt]
\hline
  \multicolumn{1}{|c|}{~} &\multicolumn{3}{|c|}{~}
    &\multicolumn{3}{c|}{~}
    &\multicolumn{3}{c|}{~} \\ [-12pt]
  $|t|$ $\left[{\rm GeV}^2\right] $ &
   \multicolumn{3}{c|}{$W=40\,{\rm GeV}$} &
   \multicolumn{3}{c|}{$W=70\,{\rm GeV}$} &
   \multicolumn{3}{c|}{$W=100\,{\rm GeV}$} \\ [3.0pt]
 \hline
 $0.10$&    $4.77$&  $\pm$ $0.50$&  $\pm$  $0.49$&    $7.81$&  $\pm$  $0.51$& $\pm$ $0.85$&   $11.0$& $\pm$ $0.85$&$\pm$ $2.23$ \\
 $0.30$&    $1.62$&  $\pm$ $0.23$&  $\pm$  $0.18$&    $2.88$&  $\pm$  $0.22$& $\pm$ $0.28$&   $3.71$& $\pm$ $0.31$&$\pm$ $0.49$ \\
 $0.50$&    $0.69$&  $\pm$ $0.11$&  $\pm$  $0.07$&    $0.91$&  $\pm$  $0.10$& $\pm$ $0.10$&   $1.18$& $\pm$ $0.13$&$\pm$ $0.16$ \\
 $0.80$&    $0.10$&  $\pm$ $0.02$&  $\pm$  $0.01$&    $0.16$&  $\pm$  $0.02$& $\pm$ $0.02$&   $0.24$& $\pm$ $0.03$&$\pm$ $0.04$ \\
 \hline 
 \hline 
  $b$ [${\rm GeV}^{-2}$]  & $5.38$ & $\pm$ $0.30$ & $\pm$ $0.23$ & $5.49$ & $\pm$ $0.19$ & $\pm$ $0.26$ & $5.49$ & $\pm$ $0.20$ &  $\pm$ $0.35$  \\
 \hline 
\end{tabular}
\caption{ The DVCS cross section $\gamma^* p\rightarrow \gamma p$, differential in $t$, $d\sigma_{DVCS}/dt$,
for three values of $Q^2$ at $W=82\,{\rm GeV}$, and 
for three values of $W$ at $Q^2=10\,{\rm GeV}^2$. Results for the corresponding $t$-slope parameters $b$ are given. 
The first errors are statistical, the second systematic.}
\label{sigtq}
\end{table}

\begin{table}[htbp]
\centering
\begin{tabular}{ll}
\begin{tabular}{|r|rrc|}
 \hline 
 $\phi$  [${\rm deg.}$]& \multicolumn{3}{c|}{ $A_C(\phi)$}\\
 \hline
 $10$    &$\;  0.326 $ &$ \pm$ $ 0.086 $ &$ \pm$ $ 0.180 $ \\
 $35$    &$\;  0.119 $ &$ \pm$ $ 0.076 $ &$ \pm$ $ 0.090 $ \\
 $70$    &$\; -0.039 $ &$ \pm$ $ 0.080 $ &$ \pm$ $ 0.030 $ \\
 $110$   &$\;  0.035 $ &$ \pm$ $ 0.092 $ &$ \pm$ $ 0.028 $ \\
 $145$   &$\; -0.234 $ &$ \pm$ $ 0.079 $ &$ \pm$ $ 0.076 $ \\
 $170$   &$\; -0.210 $ &$ \pm$ $ 0.075 $ &$ \pm$ $ 0.169 $ \\
 \hline 
\end{tabular}
\end{tabular}
\caption{ 
 The DVCS  
 beam charge asymmetry $A_C(\phi)$ as a function of
$\phi$ and integrated over the kinematic range 
$6.5 < Q^2 < 80\,{\rm GeV}^2$, $30 < W < 140\,{\rm GeV}$ 
and $|t| < 1\,{\rm GeV}^2$. The first errors are statistical, the second systematic.}
\label{tabbca}
\end{table}

\vfill
\newpage

\end{document}